\documentclass[12pt]{iopart}
\pdfoutput=1

\usepackage{iopams}  
\usepackage{graphicx}
\usepackage{cite}

\begin{document}
\title[Segregation of receptor-ligand complexes in cell adhesion zones]{Segregation of receptor-ligand complexes in cell adhesion zones:
Phase diagrams and role of thermal membrane roughness}
 
\author{B R\'{o}\.{z}ycki$^{1,2}$, R Lipowsky$^1$ and T R Weikl$^1$}

\address{$^1$ Max Planck Institute of Colloids and Interfaces, Department of Theory and Bio-Systems, 14424 Potsdam, Germany} 
\address{$^2$ Laboratory of Chemical Physics, National Institute of Diabetes and
Digestive and Kidney Diseases, National Institutes of Health, Bethesda, MD
20892-0520, USA}
%\ead{thomas.weikl@mpikg.mpg.de}

\begin{abstract}
The adhesion zone of immune cells, the `immunological synapse', exhibits characteristic domains of receptor-ligand complexes. The domain formation is likely caused by a length difference of the receptor-ligand complexes, and has been investigated in experiments in which T cells adhere to supported membranes with anchored ligands. For supported membranes with two types of anchored ligands, MHCp and ICAM1, that bind to the receptors TCR and LFA1 in the cell membrane, the coexistence of domains of TCR-MHCp and LFA1-ICAM1 complexes in the cell adhesion zone has been observed for a wide range of ligand concentrations and affinities. For supported membranes with long and short ligands that bind to the same cell receptor CD2, in contrast, domain coexistence has been observed for a rather narrow ratio of ligand concentrations. In this article, we determine detailed phase diagrams for cells adhering to supported membranes with a statistical-physical model of cell adhesion. We find a characteristic difference between the adhesion scenarios in which two types of ligands in a supported membrane bind (i) to the same cell receptor  or (ii) to two different cell receptors, which helps to explain the experimental observations. Our phase diagrams fully include thermal shape fluctuations of the cell membranes on nanometer scales, which lead to a critical point for the domain formation and to a cooperative binding of the receptors and ligands.

\end{abstract}
%\pacs{1315, 9440T}
%\submitto{\JPG}
\maketitle

%%%
\section{Introduction}
%%%

Cell adhesion is mediated by the specific binding of a variety of membrane-anchored receptor and ligand molecules. In 1990, Springer suggested that the length difference of receptor-ligand complexes in the contact zone of immune cells may lead to segregation, {\em i.e.} to the formation of domains within the cell contact zone that contain receptor-ligand complexes with different lengths \cite{Springer90}. The `length' of a receptor-complex here is the intermembrane distance, or local membrane separation at the site of the complex. A length difference between receptor-ligand complexes leads to an indirect, membrane-mediated repulsion of the complexes because the membranes have to bend to compensate the mismatch, which costs bending energy. Important receptor-ligand complexes in T-cell adhesion are the TCR-MHCp complex with a length of about 13 nm \cite{Garcia96}, the CD2-CD48 complex with the same length of 13 nm \cite{Merwe95,Wang99,Milstein08}, and the LFA1-ICAM1 complex with a length of about 40 nm \cite{Dustin00}. In 1998 and 1999, the contact zone of T cells was indeed found to contain domains that either contain the short TCR-MHCp or the long LFA1-ICAM1 complexes \cite{Monks98,Grakoui99}. As expected from their length, the CD2-CD48 complexes are located within the TCR-MHCp domains \cite{Milstein08}. However, the question whether the domain formation is predominantly caused by the length mismatch of receptor-ligand complexes is complicated by the role of the actin cytoskeleton, which polarizes during T-cell adhesion and transports clusters of TCR-MHCp complexes towards the center of the cell contact zone \cite{Mossman05,Kaizuka07,DeMond08}, and by additional, direct protein-protein interactions \cite{Douglass05}.  The domain formation is closely linked to T-cell activation, with TRC clusters forming within seconds of T-cell adhesion triggering the first activation signals \cite{Campi05,Mossman05}.

Direct evidence for a central role of the length of receptor-ligand complexes comes from experiments in which these lengths are altered by protein engineering \cite{Choudhuri05,Milstein08}. Milstein and coworkers \cite{Milstein08} have considered variants of the protein CD48 with four and five immunoglobolin-like (Ig-like) domains. The CD48 variants are longer than the  CD48 wildtype, which contains only two Ig-like domains. The CD48  wildtype and both CD48 variants bind to CD2 on T cells.  From electron micrographs of the contact zone between T cells and supported membranes that contain one of the three CD48 types, Milstein and coworkers found that the length of the CD2-CD48 complex is 12.8 $\pm$ 1.4 nm for wildtype CD48, 14.2 $\pm$ 1.2 nm for the CD48 variant with four Ig-like domains, and 15.6 $\pm$ 1.4 nm for the variant with five Ig-like domains. In fluorescence experiments of T cells on supported membranes that contain mixtures of two of the three CD48 types, Milstein and coworkers observed that CD2-CD48 wildtype complexes segregate from both CD2-CD48 variant complexes. The segregation seems to be driven by the length difference of the complexes since the T-cell cytoskeleton can only `act on' CD2 and, thus, can hardly `discriminate' between the different complexes.  However, Milstein and coworkers observe domain coexistence in the contact zone only within a narrow range of concentration ratios of CD48 wildtype and CD48 variants. For T cells adhering to supported membranes with MHCp and ICAM1, in contrast, domain coexistence has been observed for a rather wide range of MHCp and ICAM1 concentrations and affinities \cite{Grakoui99,Hailman02,Mossman05}. 
 
In this article, we calculate detailed phase diagrams for cells adhering to supported membranes with anchored ligands. We consider two general adhesion scenarios: In the first scenario, long and short ligands in the supported membrane bind to the same cell receptor (see section 4), as in the experiments of Milstein and coworkers \cite{Milstein08}, in which CD48 wildtype and a CD28 variant in the supported membrane both bind to CD2 in the T-cell membrane. In the second scenario, two types of ligands in the supported membrane bind to two types of receptors in the cell membrane (see section 5), as in experiments in which MHCp and ICAM1 in the supported membrane bind to TCR and LFA1 in the T-cell membrane. We find a characteristic difference between the phase diagrams in the two scenarios (see fig.~\ref{figure_phasediagrams}). In the first scenario, domain coexistence only occurs along a coexistence line. In the second scenario, in contrast, domain coexistence occurs in a wide coexistence region. Our phase diagrams thus help to understand why Milstein and coworkers observe domain coexistence only within a narrow range of concentration ratios.
 
Our calculations are based on a statistical-physical model of cell adhesion (see section 2). In this model, the membranes are described as elastic sheets discretized into small patches that can contain single receptor or ligand molecules \cite{Lipowsky96,Weikl01,Weikl06,Weikl09}. 
The binding and domain formation of receptor-ligand complexes is affected by thermally excited shape fluctuations of the membranes on nanometer scales. These shape fluctuations lead to a critical point for the segregation of long and short receptor-ligand complexes \cite{Asfaw06,Weikl09}. The critical point depends on the length difference of the complexes, on the concentrations and affinities of the receptors and ligands, and on the bending rigidity of the membranes (see section 4). The critical point constitutes a threshold for segregation, or domain formation, and may help to understand why Milstein and coworkers have observed segregation of wildtype CD48 from each of the two CD48 variants, but not segregation of the two CD48 variants \cite{Milstein08}. In addition, the membrane shape fluctuations on nanoscales lead to a cooperative binding of receptor-ligand complexes \cite{Krobath09} (see section 3).

%%%
\section{Statistical-physical description of cell adhesion}
%%%

\begin{figure}[t]
\hspace*{2.3cm}
\includegraphics[width = 0.65 \linewidth]{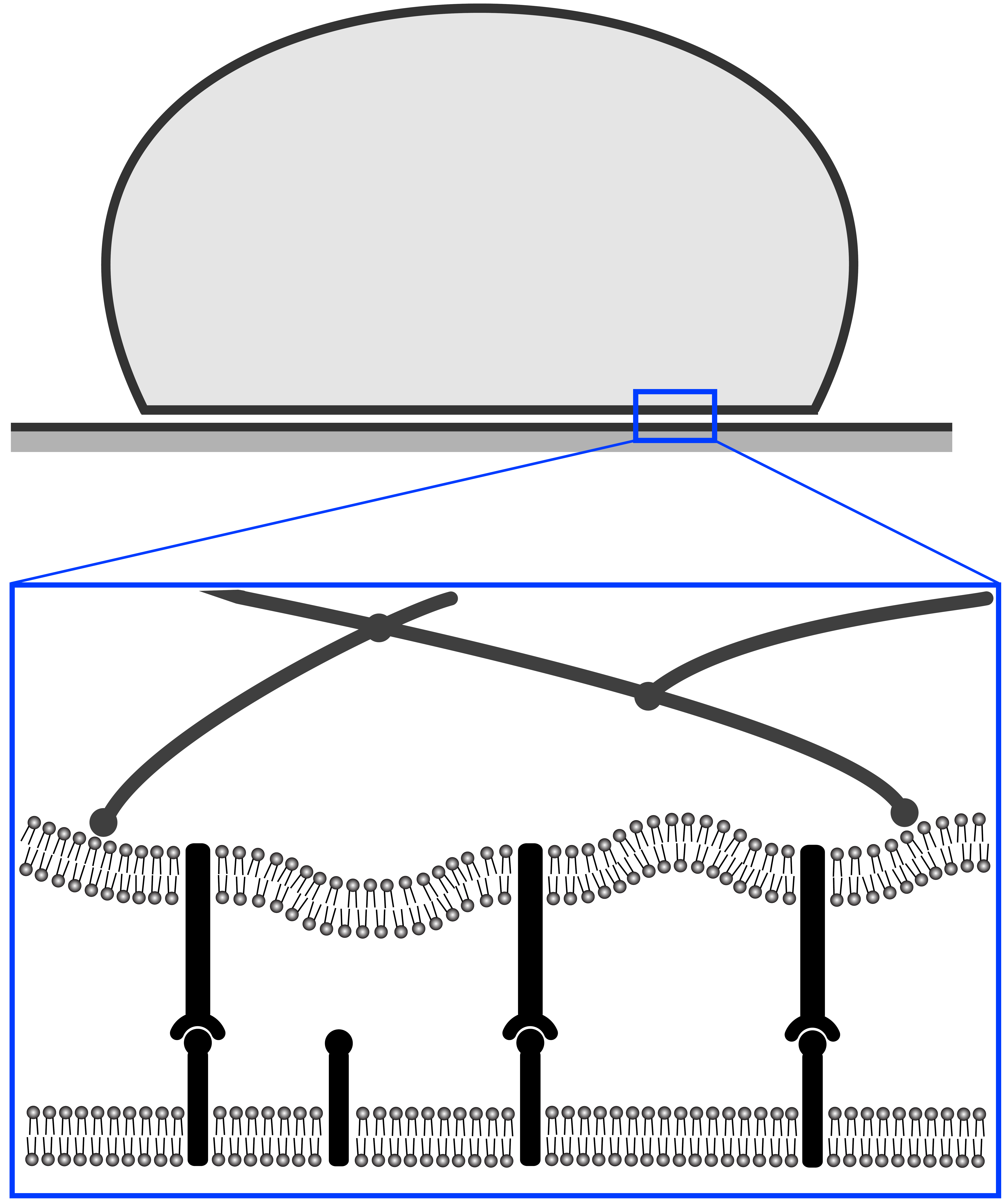}
\caption{A cell adhering to a supported membrane with anchored ligands that bind to receptors in the cell membrane. The binding of receptors and ligands in the cell adhesion zone is affected by membrane shape deformations and fluctuations on nanometer scales, which are dominated by the bending rigidity of the cell membrane. The immune cell receptors are typically mobile along the membrane and not, or only weakly \cite{DeMond08}, coupled to the cytoskeleton.
}
\label{figure_cell_adhesion}
\end{figure}

Cell adhesion involves length scales that differ by orders of magnitude (see fig.~\ref{figure_cell_adhesion}). The diameters of the cell and cell contact zone have values of several micrometers, while the average separation of the membranes within the contact zone is typically tens of nanometers. Other important length scales in the cell contact zone are the average distance between receptor-ligand bonds, and the binding width of receptor and ligand molecules. The binding width is the difference between the smallest and the largest local membrane separation at which the molecules can bind. The binding width of the typically rather stiff receptor and ligand proteins that mediate cell adhesion is much smaller than the length of the proteins. 

The binding equilibrium and segregation of receptor-ligand complexes in cell contact zones is affected by membrane shape deformations and fluctuations. Since bound receptor-ligand complexes constrain the local separation of the membranes, the relevant deformations and fluctuations of the membranes occur on lateral length scales smaller than the average distance between neighboring pairs of complexes, which is about 100 nanometers for complex concentrations of about 100 per square micrometer  \cite{Grakoui99}. It is reasonable to assume that the elasticity of the membranes is dominated by their bending rigidity on these length scales. The binding rigidity $\kappa$ dominates over the membrane tension $\sigma$ on lateral length scales smaller than the crossover length $\sqrt{\kappa/\sigma}$ \cite{Lipowsky95}, which is of the order of several hundred nanometers for cell membranes \cite{Krobath07}. The cytoskeletal elasticity \cite{Gov03,Fournier04,Lin04,Auth07} contributes on length scales larger than the average distance between the cytoskeletal anchors in the membrane, which may be around 100 nanometers \cite{Alberts02}. The bending rigidity thus is likely to dominate over the lateral tension and the cytoskeletal elasticity on lateral length scales up to 100 nanometers relevant here. 

We have developed discrete models for the adhesion of membranes {\em via} anchored receptors and ligands \cite{Lipowsky96,Weikl00,Weikl06,Weikl09}. In discrete models, the two apposing membranes in the contact zone of cells or vesicles are divided into small patches \cite{Lipowsky96,Weikl00,Weikl01,Weikl02b,Smith05,Asfaw06,Weikl06,Rozycki06a,Krobath07,Tsourkas07,Tsourkas08,Reister08,Asfaw09}. In our models, the rigidity-dominated elasticity of the membranes in the contact zones of cells or vesicles is described by \cite{Lipowsky96,Weikl06}
\begin{equation}
\mathcal{H}_{\rm el} \{ l \} = \frac{\kappa}{2 a^2} \sum_i \left( \Delta_{\rm d} l_i \right)^2
\label{elastic_energy}
\end{equation}
where $l_i$ is the local separation of the apposing membrane patches $i$.  The elastic energy depends on the mean curvature $(\Delta_{d}l_i)/a^2$ of the separation field $l_i$ with the discretized Laplacian $\Delta_{d}l_i =l_{i1}+l_{i2}+l_{i3}+l_{i4}-4 l_i$. Here $l_{i1}$ to $l_{i4}$ are the membrane separations at the four nearest-neighbor patches of membrane patch $i$ on the quadratic array of patches. The linear size $a$ of the membrane patches is chosen to be around 5 nm to capture the whole spectrum of bending deformations of the lipid membranes \cite{Goetz99}. 
The `effective bending rigidity' of the two membranes with rigidities $\kappa_1$ and $\kappa_2$ is $\kappa=\kappa_1\kappa_2/(\kappa_1 + \kappa_2)$. If one of the membranes, e.g.~membrane 2, is a planar supported membrane, the effective bending rigidity $\kappa$ equals the rigidity $\kappa_1$ of the apposing membrane since the rigidity $\kappa_2$ of the supported membrane is taken to be much larger than $\kappa_1$.

\begin{figure}[t]
\hspace*{2.3cm}
\includegraphics[width = 0.65 \linewidth]{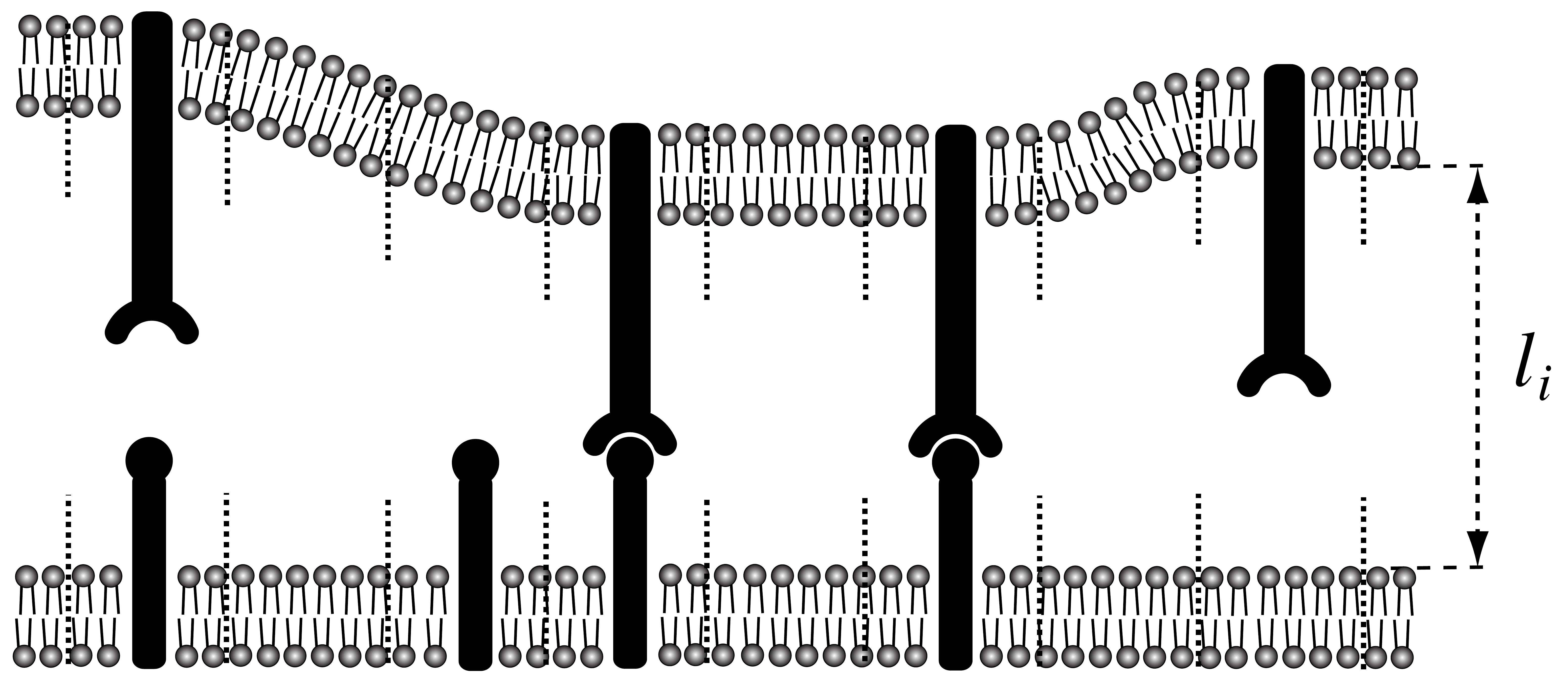}
\caption{A supported membrane with anchored ligands (bottom) that bind to receptors in an apposing cell or vesicle membrane (top). In our model, the membranes are discretized in small patches, which can contain single receptor or ligand molecules. The shape and thermal fluctuations of the cell or vesicle membrane in the  adhesion zone are described by the local separations $l_i$ of apposing membrane patches $i$. A receptor can bind to a ligand molecule (i) if the ligand is located in the membrane patch apposing the receptor, and (ii) if the local membrane separation $l_i$ is close to the length of the receptor-ligand complex.}
\label{figure_model_one}
\end{figure}

The overall energy of the membranes in the cell contact zone
\begin{equation}
\mathcal{H} \{ l, n, m \} = \mathcal{H}_{\rm el} \{ l \} + \mathcal{H}_{\rm int} \{ l, n, m \}
\end{equation}
is the sum of the elastic energy $\mathcal{H}_{\rm el} \{ l \}$ and interaction energy $\mathcal{H}_{\rm int} \{ l, n, m \}$. The interaction energy depends on the distribution $n$ of the receptors in membrane 1, on the distribution $m$ of the receptors in membrane 2, and on the separation field $l$ of the membranes. In our models, each patch of the discrete membranes can only be occupied by one receptor or ligand molecule.
Mobile receptor and ligand molecules diffuse by `hopping' from patch to patch, and the thermal fluctuations of the membranes are reflected in variations of the local separation of apposing membrane patches. A receptor can bind to a ligand molecule if the ligand is located in the membrane patch apposing the receptor, and if the local separation of the membranes is close to the length of the receptor-ligand complex (see fig.~\ref{figure_model_one} and below). In discrete models, the receptor and ligand molecules are taken into account as individual molecules. In continuum models, in contrast, the distributions of receptor and ligand molecules on the membranes are described by continuous concentration profiles \cite{Bell78,Bell84,Komura00,Bruinsma00,Chen03,Coombs04,Shenoy05,Wu06,Zhang08,Xu09,Atilgan09}.

%%%
\section{Adhesion via a single type of receptor-ligand complexes}
%%%

%%
\subsection{Interaction energy of receptors and ligands}

We first consider the case in which the adhesion is mediated by a single type of receptor-ligand complexes. Examples of this case are (i) cells adhering to supported membranes that contain a single type of ligand \cite{Dustin96,Zhu07,Milstein08,Tolentino08}, and (ii) vesicles with anchored receptors that adhere to supported membranes or surfaces with complementary ligands \cite{Albersdoerfer97,Kloboucek99,Maier01,Smith06,Lorz07,Purrucker07,Smith08,Reister08,Fenz09,Monzel09,Streicher09,Smith10}. 
The interactions of receptors and ligands within the contact zone of the cell or vesicle are described by the interaction energy \cite{Weikl01,Krobath09}
\begin{equation}
\mathcal{H}_{\rm int} \{ l, n, m \} = \sum_{i}  n_i m_i V(l_i)
\label{interaction_energy_one}
\end{equation}
in our model. Here, the occupation number $n_i=1$ or 0 indicates whether a receptor is present or absent in membrane patch $i$ of the cell, and $m_i=1$ or 0 indicates whether a ligand is present or absent in patch $i$ of the apposing membrane. Receptor and ligand molecules in apposing patches $i$ of the membranes interact with the potential $V(l_i)$.
For simplicity, we describe this interaction by the square-well potential
\begin{eqnarray}
V(l_i) &=& -U \mbox{~~for~~} l_o- l_{\rm we}/2 < l_i  < l_o + l_{\rm we}/2 \nonumber\\
  &=& 0 \mbox{~~otherwise} 
\label{square_well}
\end{eqnarray}
which depends on the binding energy $U>0$, and the length $l_o$ and binding width $l_{\rm we}$ of a receptor-ligand complex. A receptor thus binds to an apposing ligand with energy $-U$ if the local separation $l_i$ of the membranes is within the binding range $l_o \pm l_{\rm we}/2$.

\subsection{Effective adhesion potential}
\label{effective_potential_one}

The binding equilibrium of the membranes in the contact zone can be determined from the free energy  $\mathcal{F} = -k_B T \ln \mathcal{Z} $, where $\mathcal{Z}$ is the partition function of the system, $k_B$ is Boltzmann's constant, and $T$ is the temperature. The partition function $\mathcal{Z}$ is the sum over all possible membrane configurations, with each configuration $\{ l, n, m \}$ weighted by the Boltzmann factor $\exp \left[ - \mathcal{H} \{ l, n, m \} /k_B T \right]$. A membrane configuration in the contact zone is specified by the separation field $l$ of the membranes, the distribution $m$ of the receptors in the cell membrane, and the distribution $n$ of ligands in the apposing membrane. In our model, the partial summation in the partition function $\mathcal{Z}$ over all possible distributions $m$ and $n$ of receptors and ligands can be performed exactly, which leads to an effective adhesion potential. The effective adhesion potential $V_{\rm ef}(l_i)$ is a square-well potential  with (i) the same binding range $l_{\rm we}$ as the receptor-ligand interaction (\ref{square_well}) and (ii) an effective potential depth $U_{\rm ef}$ that depends on the concentrations and binding energy $U$ of receptors and ligands \cite{Weikl01,Weikl06,Krobath09}: 
\begin{eqnarray}
V_{\rm ef}(l_i) &=& -U_{\rm ef} \mbox{~~for~~} l_o- l_{\rm we}/2 < l_i  < l_o + l_{\rm we}/2 \nonumber\\
  &=& 0 \mbox{~~otherwise} 
\label{square_well}
\end{eqnarray}
For typical concentrations of receptors and ligands in cell adhesion zones up to hundred or several hundred molecules per square micrometer, the average distance between neighboring pairs of receptor and ligand molecules is much smaller than the width of the molecules. For these small concentrations, the effective binding energy of the membranes is \cite{Krobath09}
\begin{equation}
U_{\rm ef} \approx k_B T \, a^2  e^{U/k_BT}\, [R] [L ]
\label{Uef_one}
\end{equation}
where $[R]$ is the area concentration of unbound receptors in the cell membrane, and  $[L]$ is the area concentration of unbound ligands in the apposing membrane. The binding equilibrium in the contact zone thus can be determined from considering two membranes with the elastic energy (\ref{elastic_energy}) that interact {\em via} an effective adhesion potential with well depth $U_{\rm ef}$ and width $l_{\rm we}$.

\subsection{Area fraction $P_b$ of the membranes within binding range of receptors and ligands}
\label{section_Pb}

Receptor-ligand complexes can only form at membrane patches with a local separation within the binding range $l_o \pm l_{\rm we}/2$ of the receptors and ligands (see eq.~(\ref{square_well})). The area concentration $[RL]$ of the receptor-ligand complexes in the contact zone therefore is proportional to the fraction $P_b$ of these membrane patches \cite{Krobath09}:
\begin{equation}
[RL] \approx P_b\, K\, [R][L]
\label{RL}
\end{equation}
Here, $K$ is the equilibrium constant for receptor-ligand binding within this membrane fraction. In our model, the equilibrium constant is $K = a^2 e^{U/k_BT}$.

In equilibrium, the fraction $P_b$ of membrane patches with a local separation within receptor-ligand binding range depends on the effective binding energy $U_{\rm ef}$, the binding width $l_{\rm we}$, the effective rigidity $\kappa$ of the membranes, and the temperature $T$. We have found that the effect of these four quantities on $P_b$ can be captured by a single dimensionless quantity, the rescaled effective potential depth \cite{Krobath09}
\begin{equation}
u \equiv U_{\rm ef}\,\kappa\, l_{\rm we}^2 / (k_B T)^2 \approx (\kappa/k_B T) l_{\rm we}^2 K [R][L]
\end{equation}
To a first approximation, the membrane fraction $P_b$ depends only on $u$ for typical lengths and concentrations of receptor-ligand complexes in cell adhesion zones. In cell adhesion zones, direct contacts between the membranes can be neglected since the average separation of the membranes, which depends on the length $l_o$ of the complexes, is typically larger than the thermal membrane roughness \cite{Krobath09}.

\begin{figure}[h]
\hspace*{2.3cm}
\includegraphics[width = 0.6 \linewidth]{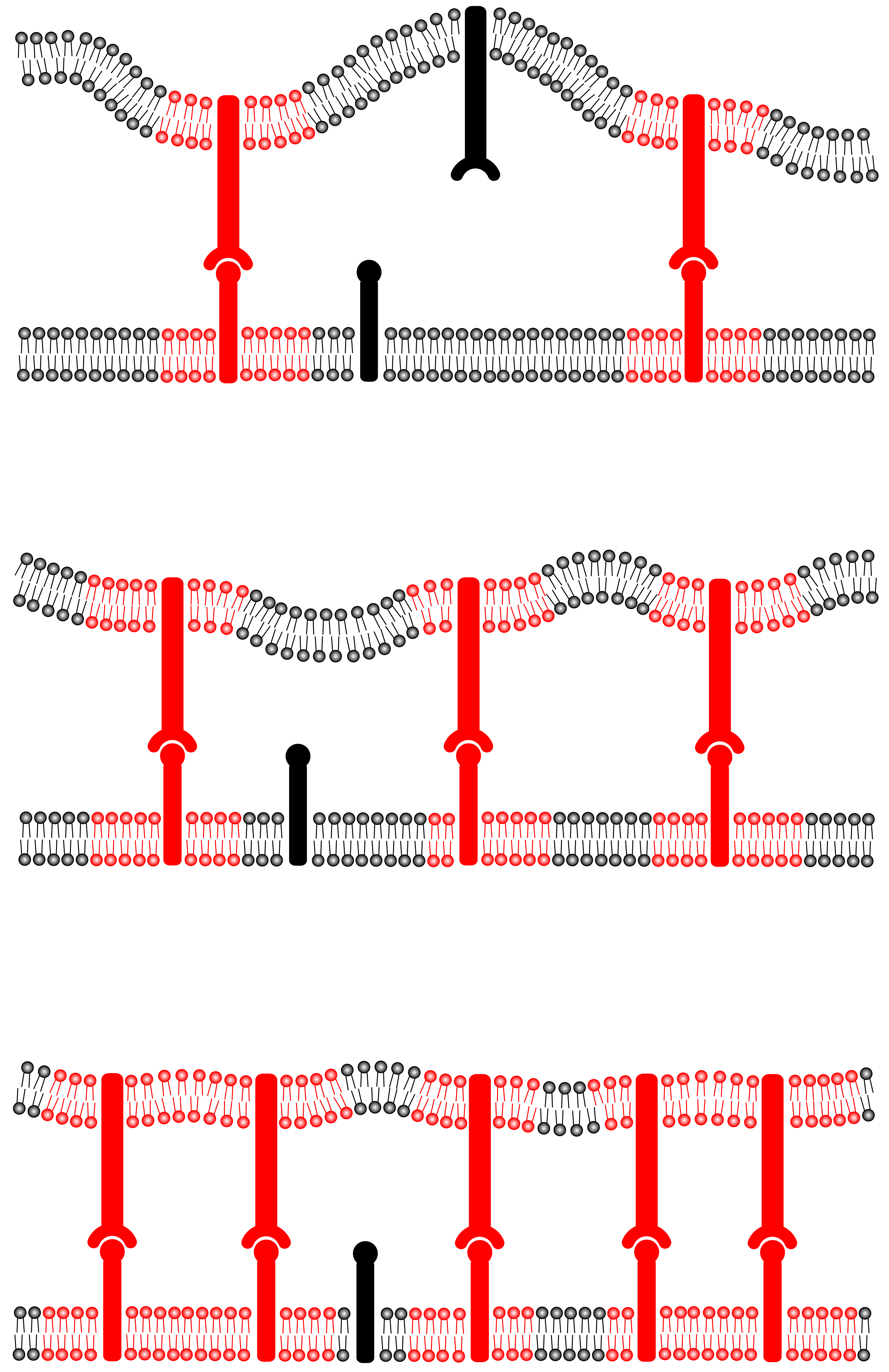}
\caption{An important quantity is the area fraction $P_b$ of the membranes within binding separation of the receptors and ligands. The area fraction $P_b$ (shown in red) increases with the concentrations of receptors and ligands, since the formation of receptor-ligand bonds `smoothens out' thermal membrane shape fluctuations. The `smoothening' facilitates the formation of additional receptor-ligand bonds and, thus, leads to a binding cooperativity \cite{Krobath09}.}
\label{figure_Pb}
\end{figure}

From Monte Carlo simulations, we have found that the functional dependence of the area fraction $P_b$ on the rescaled potential depth $u$ is well described by 
\begin{equation}
P_b \approx \frac{u}{c_1 + u} 
\label{single_para_fit}
\end{equation}
with the dimensionless coefficient $c_1 \simeq 0.071$ \cite{Krobath09}. The membrane fraction $P_b$ increases with $u$ and, thus, increases with the effective binding energy $U_{\rm ef}$ and the effective bending rigidity $\kappa$. The reason for this increase is that the roughness of the membranes resulting from thermal shape fluctuations decreases with $U_{\rm ef}$ and $\kappa$. The membrane fraction $P_b$ decreases with the temperature $T$ since the roughness increases with $T$.  The thermal roughness, defined as the standard deviation of the local membrane separation from its average, is the characteristic length scale for membrane excursions in the perpendicular direction. The membrane fraction $P_b$ within receptor-ligand binding range is much smaller than 1 if the roughness is large compared to the binding width $l_{\rm we}$ of the complexes, and close to 1 if the roughness is small compared to $l_{\rm we}$.

\subsection{Concentrations of bound and unbound receptors of an adhering cell}
\label{section_bound_unbound}

From eqs.~(\ref{RL}) to (\ref{single_para_fit}), we obtain the relation \cite{Krobath09}
\begin{equation}
[RL] \approx \frac{\kappa l_{\rm we}^2 K^2 [R]^2 [L]^2 }{c_1 k_B T + \kappa l_{\rm we}^2 K [R] [L] }
\label{RLfull}
\end{equation}
between the area concentration $[RL]$ of bound receptor-ligand complexes in the contact zone and the area concentrations $[R]$ and $[L]$ of unbound receptors and ligands. This nonlinear relation reflects the cooperative binding of receptors and ligands. This cooperativity arises because the binding of receptors and ligands suppresses thermal membrane fluctuations and, thus, smoothens the membranes, which facilitates the binding of additional receptors and ligands (see fig.~\ref{figure_Pb}). 

The total number $N$ of receptors in the cell membrane is constant. The concentrations of bound and unbound receptors are therefore connected by  the additional relation \cite{Weikl09}
\begin{equation}
N \approx [R] A + [RL] A_c
\label{Nsingle}
\end{equation}
where $A$ is the total area of the cell membrane, and $A_c$ the contact area. We have neglected here the area occupied by bound receptor-ligand complexes since this area is small compared to the total contact area $A_c$ for typical concentrations in cell adhesion zones. The concentrations of unbound receptors within and outside of the contact area then are equal. Together, the two relations (\ref{RLfull}) and (\ref{Nsingle}) determine the concentration $[R]$ of unbound receptors and the concentration $[RL]$ of bound receptors in the contact zone.

%%%
\section{Two types of membrane-anchored ligands adhering to the same cell receptor}
%%%

%%
\subsection{Interaction energy of receptors and ligands}

In recent experiments by Milstein and coworkers \cite{Milstein08}, long and short ligands anchored to a supported membrane bind to the same receptor of an adhering T cell. These ligands are wildtype CD48 and elongated CD48 variants, and the receptor in the T cell membrane is CD2. In our model, this situation is described by the interaction energy
\begin{equation}
\mathcal{H}_{\rm int} \{ l, n, m \} = \sum_i n_i \big(\delta_{m_i,1} V_1(l_i) + \delta_{m_i,2} V_2(l_i)\big) \nonumber \\
\end{equation}
for the two apposing membranes in the cell contact zone. Here, the occupation number $m_i=1$, 2, or 0 indicates whether a ligand $L_1$ of type 1, a ligand $L_2$ of type 2, or no ligand is present in patch $i$ of the supported membrane, and $n_i=1$ or 0 indicates whether a receptor $R$ is present or not in the apposing patch $i$ of the cell membrane. The Kronecker symbol $\delta_{i,j}$ equals 1 for $i=j$ and is equal to 0 for  $i\neq j$. The potential $V_1$ thus describes the interaction of the receptor $R$ with the ligand protein $L_1$ , and the potential $V_2$ the interaction between $R$ and $L_2$. For simplicity, $V_1$ and $V_2$ are again taken to be 
\begin{eqnarray}
V_1(l_i) &=& U_1 \mbox{~~for~~} l_1- l_{\rm we}/2 < l_i  < l_1+ l_{\rm we}/2 \nonumber\\
  &=& 0 \mbox{~~otherwise} 
\label{V1}
\end{eqnarray}
and 
\begin{eqnarray}
V_2(l_i) &=& U_2 \mbox{~~for~~} l_2- l_{\rm we}/2 < l_i  < l_2+ l_{\rm we}/2 \nonumber\\
  &=& 0 \mbox{~~otherwise} 
\label{V2}
\end{eqnarray}
with binding energies $U_1$ and $U_2$ and equilibrium lengths $l_1 < l_2$ of the complexes $RL_1$ and $RL_2$. We have assumed here that the two complexes have the same binding width $l_{\rm we}$.

\subsection{Effective adhesion potential}
\label{section_effective_potential_two}

\begin{figure}[t]
\hspace*{2.3cm}
\includegraphics[width = 0.65 \linewidth]{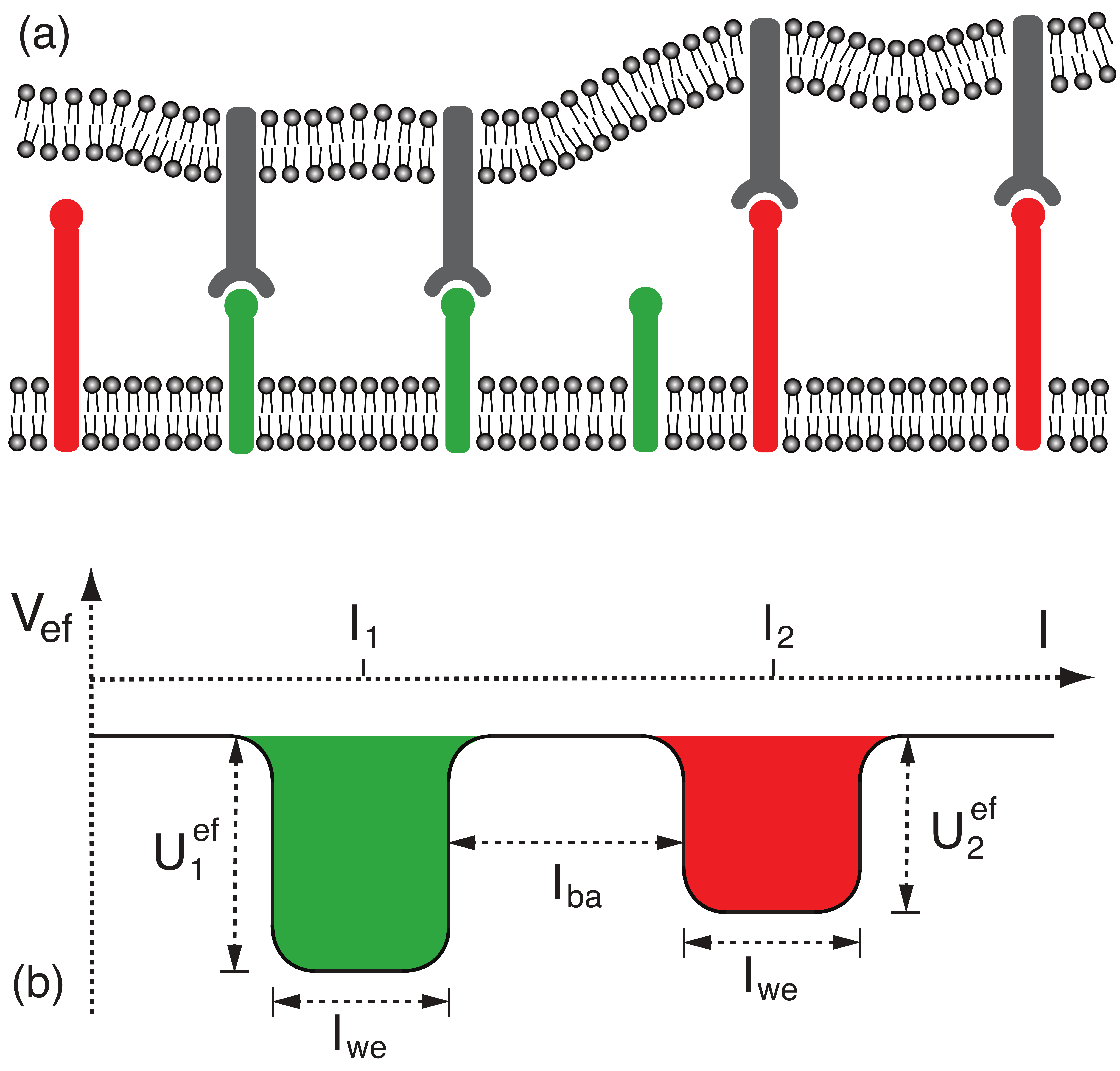}
\caption{(a) A supported membrane with short (green) ligands $L_1$  and long (red) ligands $L_2$ that bind to the same receptor $R$ in the cell membrane (top).  -- (b) The interactions of the receptors and ligands lead to an effective double-well adhesion potential $V_{\rm ef}$ of the membranes. The potential well 1 at small membrane separations $l$ reflects the interactions of the receptors with the short ligands, and the potential well 2 at larger membrane separations the interactions with the long ligands. The depths $U_1^{\rm ef}$ and $U_2^{\rm ef}$ of the two potential wells depend on the concentrations and binding constants of the receptors and ligands (see eqs.~(\ref{U1ef}) and (\ref{U2ef})). The wells have the same width $l_{\rm we}$ as the receptor-ligand interactions (\ref{V1}) and (\ref{V2}), and a separation $l_{\rm ba} = l_2 - l_1 + l_{\rm we}$ that depends on the difference between the equilibrium lengths $l_1$ and $l_2$ of the complexes $RL_1$ and $RL_2$.}
\label{figure_effective_potential}
\end{figure}

As in section \ref{effective_potential_one}, the summations over all possible distributions $m$ and $n$ of receptors and ligands in the partition function of the model leads to an effective adhesion potential \cite{Weikl09,Asfaw06}. The effective adhesion potential now is a double-well potential (see fig.~\ref{figure_effective_potential}). Both wells have the same width $l_{\rm we}$ as the potentials (\ref{V1}) and (\ref{V2}). The well with its center at the membrane separation $l_i = l_1$ reflects the interactions of the receptors $R$ with the shorter ligands $L_1$, and the well centered at $l_i = l_2$ reflects the interactions of the receptors and the longer ligands $L_2$. In analogy to eq.~(\ref{Uef_one}), the depth of the two wells 
\begin{equation}
U_{1}^{\rm ef} \approx k_B T \, K_{1}\, [R] [L_1] 
\label{U1ef}
\end{equation}
and
\begin{equation}
U_{2}^{\rm ef} \approx k_B T \, K_{2} \, [R] [L_2] 
\label{U2ef}
\end{equation}
depend on the concentrations $[R]$, $[L_1]$ and $[L_2]$ of unbound receptors and ligands, and on the binding constants $K_{1} = a^2 e^{U_1 / k_B T}$ and $K_{2} = a^2 e^{U_2 / k_B T}$ for receptors and ligands within the appropriate binding ranges \cite{Weikl09,Asfaw06}. The binding equilibrium of the receptors $R$ and ligands $L_1$ and $L_2$ in the contact zone thus can be determined from considering two apposing membranes with elastic energy (\ref{elastic_energy}) that interact {\em via} an effective double-well potential with well depths $U_{1}^{\rm ef}$ and $U_{2}^{\rm ef}$ given by eqs.~(\ref{U1ef}) and (\ref{U2ef}). 

\subsection{Phase diagram}
\label{section_phasediagram_one}

If the two wells of the effective adhesion potential are relatively shallow, thermal membrane fluctuations can easily drive membrane segments to cross from one well to the other. If the two wells are deep, the crossing of membrane segments from one well to the other well is hindered by the potential barrier of width $l_{\rm ba}$ between the wells (see fig.~\ref{figure_effective_potential}). The potential barrier induces a line tension between adjacent membrane segments that are bound in different wells \cite{Lipowsky94}. Beyond a critical depth of the potential wells, the line tension leads to the formation of large membrane domains that are bound in well one or well two. Within each domain, the adhesion of the membranes is predominantly mediated either by the receptor-ligand complexes $RL_1$ or by the complexes $RL_2$. 

We have previously found that the critical potential depth for domain formation is
\begin{equation}
U_c^{\rm ef} \approx \frac{c (k_BT)^2}{\kappa l_{\rm we} l_{\rm ba}} 
\label{Uc}
\end{equation}
with the prefactor $c = 0.225 \pm 0.02$ determined by Monte Carlo simulations \cite{Asfaw06}. Domain formation in the contact zone or, in other words, segregation of the complexes $RL_1$ and $RL_2$ can only occur if the effective potential depths $U_{1}^{\rm ef}$ and $U_{2}^{\rm ef}$ exceed the critical potential depth $U_c^{\rm ef}$. The critical potential depth depends on the temperature $T$ and the bending rigidity $\kappa$ as well as on the width $l_{\rm we}$ and separation $l_{\rm ba}$ of the two potential wells. In deriving eq.~(\ref{Uc}), we have neglected direct membrane-membrane contacts, which is reasonable for typical concentrations and lengths of receptor-ligand complexes in cell adhesion zones \cite{Asfaw06,Krobath09}. For these complex concentrations and lengths, the thermal membrane roughness is smaller than the lengths of the receptor-ligand complexes.

\begin{figure}[t]
\hspace*{2.3cm}
\includegraphics[width =  0.98 \linewidth]{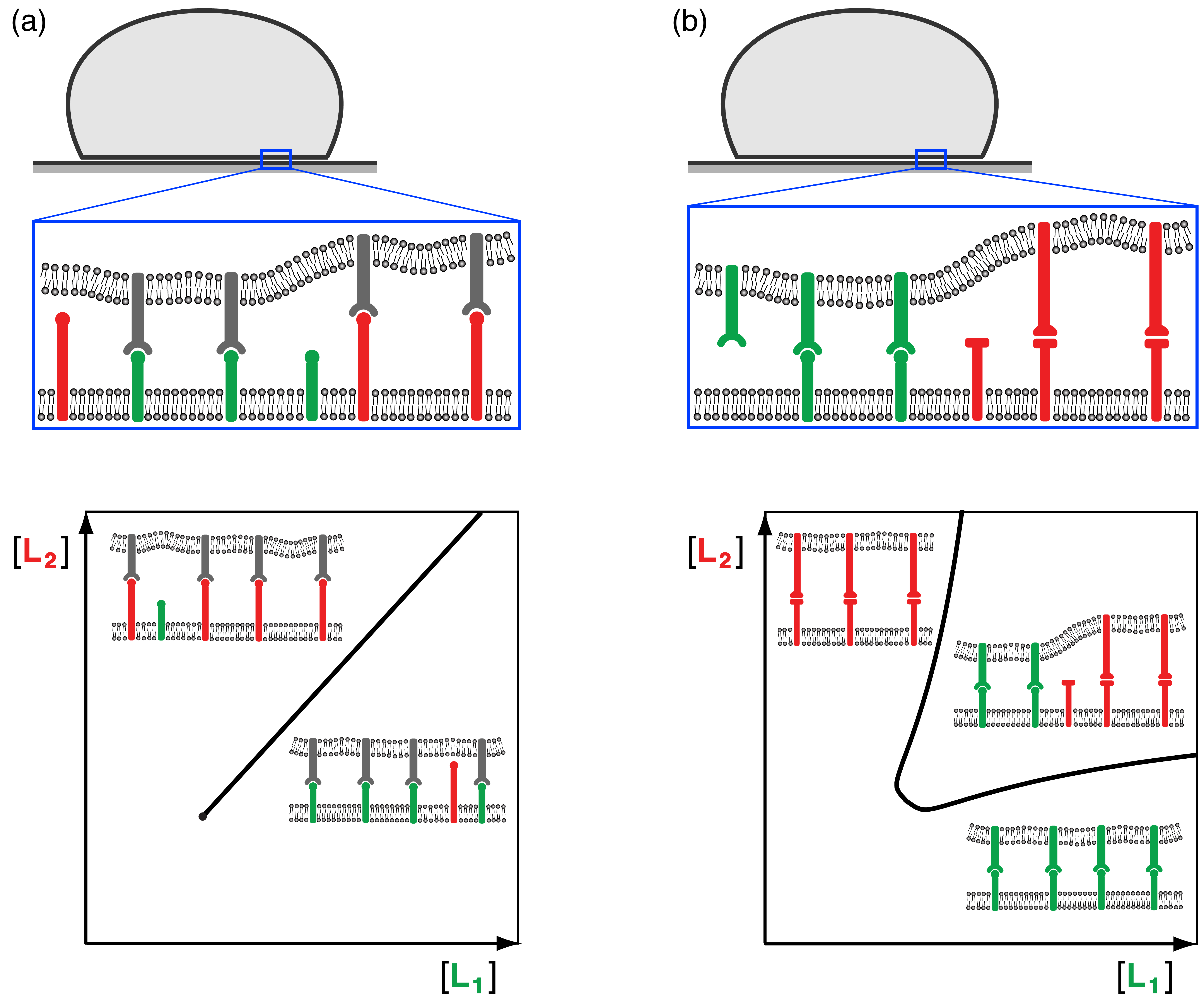}
\caption{(a) A supported membrane with two types of ligands, $L_1$ and $L_2$, that bind to the same receptor $R$ of an adhering cell (top). The different length of the ligands causes a membrane-mediated repulsion between the receptor-ligand complexes $RL_1$ and $RL_2$. For sufficiently large concentrations $[L_1]$ and $[L_2]$ of the ligands, the repulsion of the complexes $RL_1$ and $RL_2$ leads to the formation of domains. However, domain coexistence in the cell adhesion zone only occurs for equal effective binding strengths $K_1 [L_1] = K_2 [L_2]$  of the ligands. Here, $K_1$ and $K_2$ are the binding equilibrium constants of the two ligands at appropriate membrane separations. Domain coexistence thus occurs along the shown line with slope $K_1/K_2$ in the $[L_1]$-$[L_2]$-plane (bottom). The line ends at the critical point for domain formation. For $K_1 [L_1] > K_2 [L_2]$, the adhesion is dominated by the ligand
 $L_1$ throughout the cell adhesion zone, and by the ligand $L_2$ for $K_1 [L_1] < K_2 [L_2]$. We have assumed here that the supported membrane is large compared to the adhesion zone, which implies that the concentrations $[L_1]$ and $[L_2]$ of unbound ligands do not change significantly upon adhesion. -- (b) A supported membrane with two types of ligands, $L_1$ and $L_2$, that bind to different cell receptors $R_1$ and $R_2$. Domain coexistence in the cell adhesion zone now occurs for $K_1 [R_1][L_1] = K_2 [R_2][L_2]$, which leads to a broad coexistence region in the $[L_1]$-$[L_2]$-plane since the concentrations of unbound receptors $[R_1]$ and $[R_2]$ depend on the numbers of bound receptors and, thus, on the adhesion zone fractions occupied by the two domains.  
 }
\label{figure_phasediagrams}
\end{figure}

Domain coexistence occurs for equal depths 
\begin{equation}
U^{\rm ef}_1 = U^{\rm ef}_2
\end{equation}
of the potential wells if the two wells have the same width $l_{\rm we}$ as in Fig.~\ref{figure_effective_potential}. With eqs.~(\ref{U1ef}) and (\ref{U2ef}), this coexistence condition implies that domain coexistence occurs along the line with 
\begin{equation}
K_1 [L_1] = K_2 [L_2]
\end{equation}
in the $[L_1]$-$[L_2]$-plane. The line has the slope $K_1 /K_2$ and ends at a critical point (see phase diagram in fig.~\ref{figure_phasediagrams}(a)). For $U^{\rm ef}_1 > U^{\rm ef}_2$, we have $K_1 [L_1] > K_2 [L_2]$. The adhesion is then dominated by the short complexes $RL_1$ throughout the cell contact zone. For $U^{\rm ef}_1 < U^{\rm ef}_2$, in contrast, the adhesion is dominated by the long complexes $RL_2$ in the whole  contact zone. If the supported membrane is much larger than the cell contact zone, it seems reasonable to assume that the concentrations $[L_1]$ and $[L_2]$ of unbound ligands do not change upon adhesion since the `ligand reservoir' in the supported membrane is large. The concentrations $[L_1]$ and $[L_2]$ in our model then correspond to the experimental ligand concentrations in the supported membrane prior to adhesion, and our phase diagram in fig.~\ref{figure_phasediagrams}(a) to the phase diagram in fig.~7 of Ref.~\cite{Milstein08}, see Discussion. 

%%%
\section{Two types of membrane-anchored ligands adhering to different cell receptors}
%%%

%%
\subsection{Interaction energy of receptors and ligands and effective adhesion potential}

Several experimental groups have investigated the adhesion of T cells to supported membranes with anchored MHCp and ICAM1 ligands \cite{Grakoui99,Hailman02,Mossman05,Krogsgaard05,Yokosuka05,Varma06,Yokosuka08,DeMond08,Huppa10}. The ligand MHCp binds to the T cell receptor (TCR), and the ligand ICAM1 to the integrin LFA1 in the T cell membrane. The TCR-MHCp complex has a length of around 13 nm \cite{Garcia96}, and the LFA1-ICAM1 complex a length of 40 nm \cite{Dustin00}. A situation in which two ligands in the supported membrane bind to different receptors in a cell membrane can be described in our model {\em via} the interaction energy \cite{Asfaw06}
\begin{equation}
\mathcal{H}_{\rm int} \{ l, n, m \} = \sum_i \big(\delta_{n_i,1}\delta_{m_i,1} V_1(l_i) + \delta_{n_i,2}\delta_{m_i,2} V_2(l_i)\big) \nonumber \\
\end{equation}
Here, the occupation number $n_i=1$, 2, or 0 indicates whether a receptor $R_1$, a receptor $R_2$, or no receptor is present in patch $i$ of the cell membrane in the contact zone, while  $m_i=1$, 2, or 0 indicates whether a ligand $L_1$, a ligand $L_2$, or no ligand is present in the apposing patch $i$ of the supported membrane. The interaction of a receptor $R_1$ with an apposing ligand $L_1$ is described by the potential $V_1(l_i)$, and the interaction of $R_2$ with $L_2$ by the potential $V_2(l_i)$. As in section \ref{section_effective_potential_two}, a summation over all possible distributions $n$ and $m$ of receptors and ligands in the partition function leads to an effective double-well potential of the membranes. The effective potential has the same form as in fig.~\ref{figure_effective_potential}(b), but the depths of the two wells
\begin{equation}
U_{1}^{\rm ef} \approx k_B T [R_1] [L_1] K_{1}
\label{U1ef_two}
\end{equation}
and
\begin{equation}
U_{2}^{\rm ef} \approx k_B T [R_2] [L_2] K_{2}
\label{U2ef_two}
\end{equation}
now depend on the concentrations $[R_1]$ and $[R_2]$ of unbound receptors in the cell membrane, on the concentrations $[L_1]$ and $[L_2]$ of unbound ligands in the supported membrane, and on the binding constants $K_{1} = a^2 e^{U_1 / k_B T}$ and $K_{2} = a^2 e^{U_2 / k_B T}$ for the complexes $R_1L_1$ and $R_2L_2$ \cite{Weikl09}.

\subsection{Phase diagram}

As in section \ref{section_phasediagram_one}, domain coexistence in the cell contact zone requires equal depths  $U_{1}^{\rm ef} = U_{2}^{\rm ef}$ of the potential wells if the two wells have the same width $l_{\rm we}$. The effective adhesion potential then is a symmetric double-well potential. We assume here again that the total area of the supported membrane is large compared to the cell contact zone. In this case, the numbers of bound ligands in the contact zone is negligible compared to the numbers of unbound ligands in the total supported membrane, which implies that the concentrations $[L_1]$ and $[L_2]$ of unbound ligands do not change during adhesion. However, the concentrations $[R_1]$ and $[R_2]$ of unbound receptors in general change during cell adhesion because the contact area is typically a substantial fraction of the overall area of the cell membrane, and because the total numbers $N_1$ and $N_2$ of the receptors in the cell membrane are constant. During adhesion, a smaller or larger fraction of the receptors will form bound complexes $R_1L_1$ or $R_2L_2$ (see also section \ref{section_bound_unbound}). The concentrations $[R_1L_1]$ and $[R_2L_2]$ of the receptor-ligand complexes in the cell contact area depend on the fractions $P_1$ and $P_2$ of the membranes within well 1 and well 2 of the effective adhesion potential. The receptor-ligand complexes $R_1L_1$ can only form in the membrane fraction $P_1$ of the contact area within binding range $l_1 \pm l_{\rm we}/2$ of $R_1$ and $L_1$, and the complexes $R_1L_1$ only in the membrane fraction $P_2$ within binding range $l_2 \pm l_{\rm we}/2$ of $R_2$ and $L_2$. 

\begin{figure}[t]
\hspace*{2.3cm}
\includegraphics[width = 0.55\linewidth]{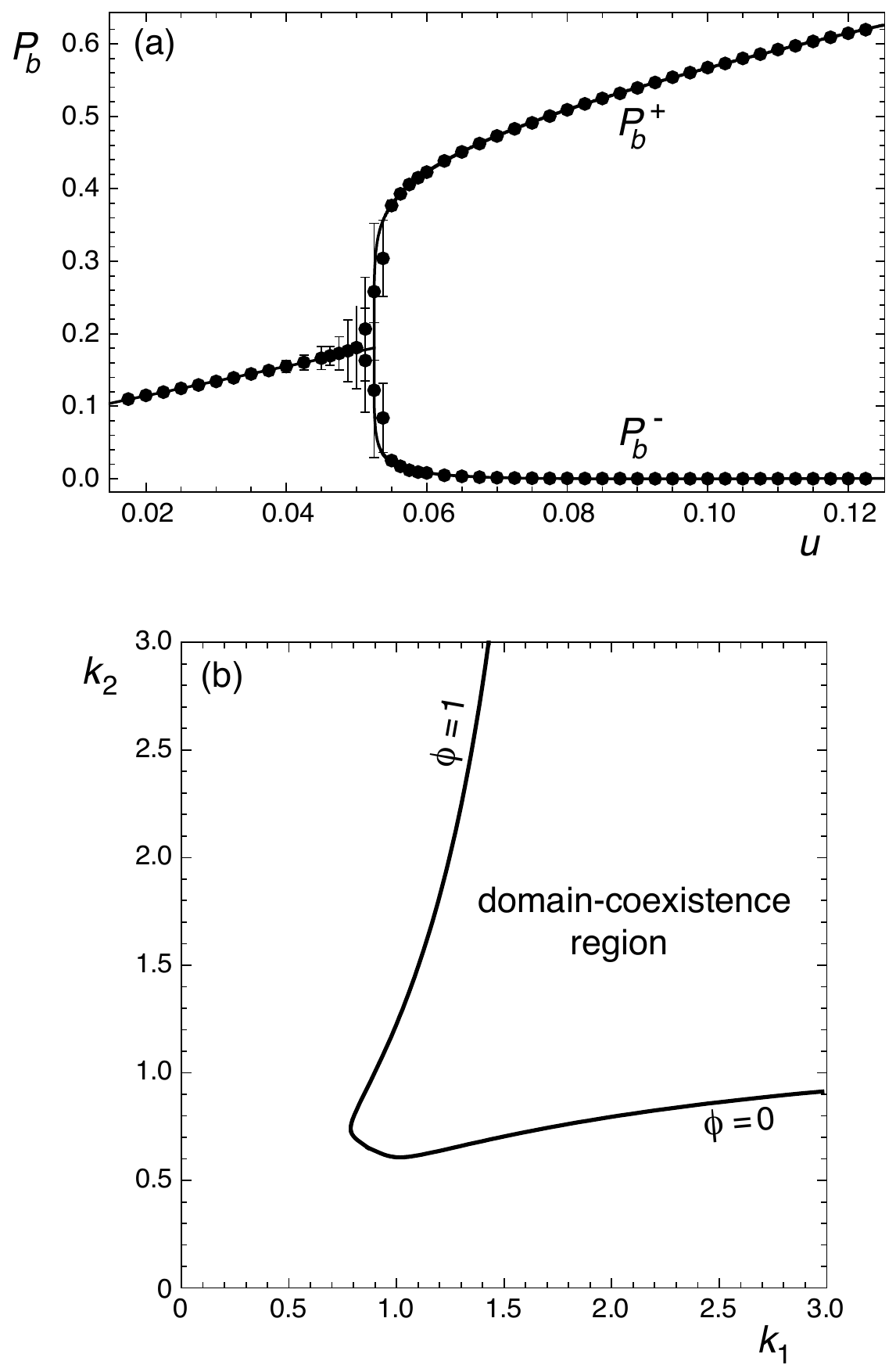}
\caption{(a) Monte Carlo data for the membrane fraction $P_b$ within the wells of a symmetric double-well potential with rescaled well depth $u$ (see eq.~(\ref{u_two})).  The data are from simulations with the rescaled width $z_{\rm we} = (l_{\rm we}/a)\sqrt{\kappa/k_BT} = 0.5$ and rescaled separation $z_{\rm ba} = (l_{\rm ba}/a)\sqrt{\kappa/k_BT} = 2$ of the two wells. Here, $l_{\rm we}$ and $l_{\rm ba}$ are the width and separation of the wells, $a$ is the linear size of the discrete membrane patches, and $\kappa$ is the effective rigidity of the membranes. Below the critical well depth $u_c \approx 0.053$, the membrane is bound in both wells with the same fraction $P_b(u)$. Above the critical well depth $u_c$, the membrane is predominantly bound in one of the two wells. The membrane fraction bound in the dominant well is $P_b^+(u)$ (upper branch for $u>u_c$), and the membrane fraction bound in the other well is $P_b^-(u)$ (lower branch for $u>u_c$). We have obtained the data from simulations with a square lattice of up to $200 \times 200$  membrane patches and with up to $5 \cdot 10^7$  Monte Carlo steps per lattice site. Details of the Monte Carlo simulations are described in \cite{Weikl06,Krobath09}. -- (b) Exemplary phase diagram for the rescaled receptor numbers $n_1 = 0.07$ and $n_2 = 0.09$ (see eq.~(\ref{n1n2})) obtained from interpolation of the Monte Carlo data in (a) and insertion of the functions $P_b^+(u)$ and $P_b^-(u)$ in eqs.~(\ref{phi=0}) and (\ref{phi=1}). Here, $k_1$ and $k_2$ are the rescaled concentrations of the two ligands in the supported membrane (see eq.~(\ref{k1k2})). The two-phase coexistence region is bounded by two lines along which the area fraction $\phi$ of domain 1 in the contact zone is $\phi=0$ and $\phi=1$ (see eqs.~(\ref{phi=0}) and (\ref{phi=1})).}
\label{figure_MC}
\end{figure}

For the symmetric double-well potential with $U_{1}^{\rm ef} = U_{2}^{\rm ef}$, the membrane fractions $P_1$ and $P_2$ within well 1 and well 2 depend primarily on the rescaled potential depth
\begin{equation}
u \equiv U_1^{\rm ef} \kappa l_{\rm we}^2/(k_B T)^2= U_2^{\rm ef} \kappa l_{\rm we}^2/(k_B T)^2
\label{u_two}
\end{equation}
as in section \ref{section_Pb}. The Monte Carlo data in fig.~\ref{figure_MC}(a) illustrate how $P_1$ and $P_2$ depend on $u$. Below the critical potential depth $u_c$, the membrane fluctuates between the two wells. Because of the symmetry of the potential, $P_1$ and $P_2$ attain the same value $P_b(u)\equiv P_1(u) = P_2(u)$ for $u<u_c$. Above the critical potential depth $u_c$, we have a spontaneous symmetry breaking of the membranes into domains that are predominantly bound in well 1 or well 2, or in other words,  predominantly bound by the complexes $R_1L_1$ or the complexes $R_2L_2$. The symmetry breaking for $u>u_c$  is reflected by two branches $P_b^{+}(u)$ and $P_b^{-}(u)$ of the membrane fraction $P_b(u)$ within the wells  (see fig.~\ref{figure_MC}(a)). For the domain predominantly bound in well 1, we have $P_1(u) = P_b^{+}(u)$ and $P_2(u) = P_b^{-}(u)$. For the domain predominantly bound in well 2, we have $P_2(u) = P_b^{+}(u)$ and $P_1(u) = P_b^{-}(u)$. For $u \gg u_c$, we have $P_b^{-}(u)\approx 0$ (see fig.~\ref{figure_MC}(a)), which implies that domain 1 then contains only the complexes $R_1L_1$, and domain 2 only the complexes $R_2L_2$.

Since the total numbers $N_1$ and $N_2$ of receptors 1 and 2 in the cell membrane are constant, we have
\begin{eqnarray}
N_1 &=& [R_1] A + K_1 [R_1][L_1]  A_c \left( \phi P_b^+(u) + (1-\phi) P_b^-(u) \right) \label{N1}\\
N_2 &=& [R_2] A + K_2 [R_2][L_2]  A_c \left( (1-\phi) P_b^+(u) + \phi P_b^-(u) \right) \label{N2}
\end{eqnarray}
Here, $A$ is the total area of the cell, $A_c$ the contact area, and $\phi$ is the fraction of the contact area occupied by domain 1, which is predominantly bound in well 1. The first terms on the right-hand sides, $[R_1]A$ and $[R_2]A$, are the total numbers of unbound receptors.
The concentrations $[R_1]$ and $[R_2]$ of unbound receptors within and outside of the contact zone are equal since we neglect the area occupied by bound receptor-ligand complexes within the contact zone (see also eq.~(\ref{Nsingle})). The second terms on the right-hand sides of eqs.~(\ref{N1}) and (\ref{N2}) are the numbers of bound receptors. In analogy to eq.~(\ref{RL}), the number of bound receptors $R_1$ in the domain predominantly bound in well 1 is $K_1 [R_1][L_1]  A_c  \phi P_b^+(u)$, and the number of bound receptors $R_2$ in this domain is $K_2 [R_2][L_2]  A_c  \phi P_b^-(u)$. The number of bound receptors $R_1$ in the domain that is predominantly bound in well 2 is $K_1 [R_1][L_1]  A_c  (1-\phi) P_b^-(u)$, and the number of bound receptors $R_2$ in this domain is $K_2 [R_2][L_2]  A_c (1- \phi) P_b^+(u)$. Below the critical potential depth $u_c$, we 
have $P_b^+(u) = P_b^-(u) = P_b(u)$. The two equations (\ref{N1}) and (\ref{N2}) therefore are independent from each other for $u<u_c$, but dependent on each other for $u>u_c$.

With the four independent, dimensionless parameters 
\begin{equation}
n_1 = \frac{N_1 \kappa l_{\rm we}^2} {A_c k_B T} ,  \quad n_2 = \frac{N_2 \kappa l_{\rm we}^2} {A_c k_B T} 
\label{n1n2}
\end{equation}
and
\begin{equation}
k_1 = K_1 [L_1] \frac{A_c}{A} ,  \quad  k_2 = K_2 [L_2] \frac{A_c}{A}
\label{k1k2}
\end{equation}
the eqs.~(\ref{N1}) and (\ref{N2}) can be rewritten as
\begin{eqnarray}
n_1 &=& \frac{u}{k_1} + u \left( \phi P_b^+(u) + (1-\phi) P_b^-(u) \right)  \label{n1}\\
n_2 &=& \frac{u}{k_2} + u \left( (1-\phi) P_b^+(u) + \phi P_b^-(u) \right)  \label{n2}
\end{eqnarray}
since we have $u  = K_1 [R_1][L_1] l_{\rm we}^2 \kappa/k_B T = K_2 [R_2][L_2] l_{\rm we}^2 \kappa/k_B T$ (see eqs.~(\ref{U1ef_two}) to (\ref{u_two})). From these two equations, one can determine $u$ and $\phi$ as functions of the independent parameters $n_1$, $n_2$, $k_1$ and $k_2$. Domain coexistence in the cell contact zone occurs for $u>u_c$ and $0<\phi<1$.

To obtain general relations for the critical point and the boundary lines of the two-phase region in the $k_1$-$k_2$ plane, we first solve eq.~(\ref{n1}) for $k_1$ and eq.~(\ref{n2}) for $k_2$, which leads to
\begin{eqnarray}
k_1 &=& \frac{u}{n_1 - u \left( \phi P_b^+(u) + (1-\phi) P_b^-(u) \right)}  \label{k1}\\
k_2 &=& \frac{u}{n_2 - u \left( (1-\phi) P_b^+(u) + \phi P_b^-(u) \right)}  \label{k2}
\end{eqnarray}
At the critical point,  we have $P_b^+(u_c) = P_b^-(u_c) = P_b(u_c)$. By inserting these relations into eqs.~(\ref{k1}) and (\ref{k2}), we obtain a general expression for the  location $(k_1,k_2)_c$ of the critical point in the $k_1$-$k_2$ plane:
\begin{equation}
\left(k_1,k_2\right)_c = \left(\frac{u_c}{n_1 - u_c P_b(u_c)}, \frac{u_c}{n_2 - u_c P_b(u_c)}\right)
\end{equation}
For the Monte Carlo data of fig.~\ref{figure_MC}(a) with $u_c \simeq 0.053$ and $P_b(u_c) \simeq 0.18$, for example, we have $\left(k_1,k_2\right)_c \simeq \left(0.053/(n_1 -0.0095), 0.053/(n_2 - 0.0095) \right)$. Since $k_1$ and $k_2$ have to be positive, domain coexistence can only occur if $n_1$ and $n_2$ are both larger than $u_c P_b(u_c)$. The domain-coexistence region  in the $k_1$-$k_2$ plane is bounded by two lines with $\phi = 0$ and $\phi = 1$. Inserting $\phi=0$ in the eqs.~(\ref{k1}) and (\ref{k2}) leads to the parametric form 
\begin{equation}
 (k_1,k_2)_{\phi=0} = \left( \frac{u}{n_1 - u P_b^-(u)},  \frac{u}{n_2 - u P_b^+(u)}\right) \mbox{~~for $u> u_c$}
 \label{phi=0}
\end{equation} 
for the $\phi=0$ line. Similarly, inserting $\phi=1$ in the  eqs.~(\ref{k1}) and (\ref{k2}) leads to the parametric form
\begin{equation}
 (k_1,k_2)_{\phi=1} = \left( \frac{u}{n_1 - u P_b^+(u)},  \frac{u}{n_2 - u P_b^-(u)}\right) \mbox{~~for $u> u_c$}
 \label{phi=1}
\end{equation} 
for the $\phi=1$ line. The domain-coexistence region in the phase diagram of fig.~\ref{figure_MC}(b), for example, follows from inserting the functions 
$P_b^+(u)$ and $P_b^-(u)$ obtained from interpolation of the Monte Carlo data shown in fig.~\ref{figure_MC}(a) into eqs.~(\ref{phi=0}) and (\ref{phi=1}).

For $u\gg u_c$, we have $P_b^-(u)\approx 0$. The $\phi=0$ line then is given by
\begin{equation}
k_2\Big|_{\phi=0} \approx \frac{k_1n_1}{n_2 - k_1n_1 P_b^+(k_1 n_1)} 
\label{k2_phi0}
\end{equation}
since we have $u \approx k_1 n_1$ for $P_b^-(u)\approx 0$.  Similarly, the $\phi=1$ line is given by 
\begin{equation}
k_1\Big|_{\phi=1} \approx \frac{k_2n_2}{n_1 - k_2n_2 P_b^+(k_2 n_2)} 
\label{k1_phi1}
\end{equation}
because of $u \approx k_2 n_2$. For $u\gg u_c$, the membranes are only bound {\em via} one of the wells. The membrane fraction $P_b^+$ bound in this well therefore can be approximated by the same expression $P_b^+(u) \approx u/(c_1+u)$ with $c_1\simeq 0.071$ as in the case of an effective single-well adhesion potential (see eq.~(\ref{single_para_fit})).

The $\phi = 0$ line has a vertical asymptote in the $k_1$-$k_2$ plane, since $k_2$ in eq.~(\ref{k2_phi0}) diverges for 
\begin{equation}
n_2 = k_1 n_1 P_{b}^{+} \left( k_1 n_1 \right)
\label{n2_asymp}
\end{equation}
because the denominator of the right-hand side of eq.~(\ref{k2})  vanishes. With $P_{b}^{+} (u) \approx u/(c_1 + u)$, we obtain the location
\begin{equation}
k_1 \approx \frac{n_2}{2 n_1} \left( 1 + \sqrt{1 + 4 \frac{c_1}{n_2} \, } \right)
\label{k1_asymptote}
\end{equation}
for this vertical asymptote from eq.~(\ref{n2_asymp}). Similarly, the $\phi=1$ line has a horizontal asymptote in the $k_1$-$k_2$ plane since the 
 the denominator of the right-hand side of eq.~(\ref{k1_phi1}) vanishes for 
\begin{equation}
n_1 = k_2 n_2 P_{b}^{+} \left( k_2 n_2 \right)
\label{n1_asymp}
\end{equation}
With $P_{b}^{+} (u) \approx u/(c_1 + u)$, we obtain the value 
\begin{equation}
k_2 \approx \frac{n_1}{2 n_2} \left( 1 + \sqrt{1 + 4 \frac{c_1}{n_1} \, } \right)
\label{k2_asymptote}
\end{equation}
for the horizontal asymptote of the $\phi=1$ line from eq.~(\ref{n1_asymp}).

%%%
\section{Discussion and Conclusions}
%%%

In this article, we have determined phase diagrams for cells adhering to supported membranes with anchored ligands. For supported membranes  with short and long ligands $L_1$ and $L_2$  that bind to the same cell receptor $R$, coexistence of domains of $L_1R$ and $L_2R$ complexes in the cell adhesion zone only occurs for equal effective binding strength $K_1[L_1] = K_2[L_2]$ of the complexes where $K_1$ and $K_2$ are the binding equilibrium constants at appropriate membrane separations. The domain coexistence thus occurs along a line in the $[L_1]$-$[L_2]$ plane, which ends at the critical point (see fig.~\ref{figure_phasediagrams}(a)). We have assumed that the area of the supported membrane is much larger than the adhesion zone, which implies that the concentrations $[L_1]$ and $[L_2]$ of unbound ligands do not change significantly during adhesion since the supported membrane constitutes a large `ligand reservoir'. Constant concentrations of unbound ligands imply constant chemical potentials $\mu_1 \approx k_B T \ln (a^2 [L_1])$ and $\mu_2 \approx k_B T \ln(a^2 [L_2])$  of the ligands in our model (see eq.~(32) in ref.~\cite{Krobath09}). A coexistence line as in the diagram of fig.~\ref{figure_phasediagrams}(a)) is typical for phase diagrams in grand-canonical ensembles with constant chemical potentials.

For supported membranes with two types of ligands $L_1$ and $L_2$ that bind to different cell receptors $R_1$ and $R_2$, we obtain a qualitatively different phase diagram with a broad coexistence region (see fig.~\ref{figure_phasediagrams}(b)). Domain coexistence in the cell adhesion zone occurs for $K_1 [R_1][L_1] = K_2 [R_2][L_2]$. The broad coexistence region is a consequence of the fact that the concentrations of unbound receptors $[R_1]$ and $[R_2]$ depend on the numbers of bound receptors and, therefore, on the fractions of the cell adhesion zone occupied by the domains of $R_1L_1$ and $R_2L_2$ complexes, since the total numbers $N_1$ and $N_2$ of receptors in the cell membrane are constant. A broad coexistence region as in the diagram of fig.~\ref{figure_phasediagrams}(b)) is typical for phase diagrams in canonical ensembles with constant particle numbers. 
 
Milstein and coworkers \cite{Milstein08} have observed domain coexistence for a narrow concentration ratio of short and long 
ligands that bind to the same cell receptor CD2, in agreement with our phase diagram in fig.~\ref{figure_phasediagrams}(a). However, two differences between our phase diagram in fig.~\ref{figure_phasediagrams}(a) and the phase diagram of Milstein and coworkers in fig.~7 of ref.~\cite{Milstein08} are: First, the coexistence line in the phase diagram of Milstein and coworkers seems to have a finite width. Such a finite width may result from slight changes of the ligand concentrations upon binding, since several cells adhere to the same supported membrane in the experiments. Second, the coexistence line in the diagram of Milstein and coworkers ends in a region in which the cells do not adhere, while the coexistence line in the diagram of fig.~\ref{figure_phasediagrams}(a) ends at a critical point. In this article, we have neglected repulsive interactions from, e.g., the cell glycocalyx. In our model, such repulsive interactions lead to an unbinding of the membranes at certain well depths $U_1^{\rm ef}$ and $U_2^{\rm ef}$ of the effective adhesion potential shown in fig.~\ref{figure_effective_potential} \cite{Asfaw06}. We obtain a phase diagram similar to the diagram of Milstein and coworkers if the well depths $U_1^{\rm ef}$ and $U_2^{\rm ef}$ at which the membranes unbind are larger than the critical potential depth $U_c^{\rm ef}$, which determines the location of the critical point in the diagram of fig.~\ref{figure_phasediagrams}(a).

We find that thermal membrane shape fluctuations on nanometer scales play a central role during cell adhesion. Fluctuations on these scales have been recently reported for immune cells adhering to coated substrates \cite{Pierres08,Pierres09}. In previous work, we have found that the fluctuations lead to a cooperative binding of receptors and ligands (see fig.~\ref{figure_Pb}) \cite{Krobath09}, and to a critical point for the segregration of long and short receptor-ligand complexes \cite{Asfaw06,Weikl09}. Our phase diagrams in fig.~\ref{figure_phasediagrams} are therefore qualitatively different from phase diagrams  calculated under neglection of shape fluctuations \cite{Coombs04}. The binding cooperativity of receptors and ligands arises since a receptor-ligand complex locally constrains the membrane shape fluctuations and facilitates the binding of nearby complexes. The binding cooperativity is thus closely related to the fluctuation-induced attractive interactions between bound receptor-ligand complexes \cite{Weikl06,Bruinsma94,Weikl00,Weikl01,Farago10}, which result from a suppression of membrane-shape fluctuations, similar to the fluctuation-induced interactions of rigid membrane inclusions \cite{Goulian93,Netz95,Golestanian96,Weikl01b}. 

We have neglected here the line tension of the domain boundaries, which may suppress the formation of small domains in the cell adhesion zone. In classical nucleation theory, the line tension leads to a threshold size for stable domains.  Experimental observations of stable microdomains in the adhesion zones of immune cells \cite{Campi05,Mossman05,Yokosuka05,Yokosuka08} indicate that this threshold size is rather small. We will consider the line tension between domains of short and long receptor-ligand complexes in detail in a future article.

\end{document}